\documentclass[aip,apl,reprint,graphicx]{revtex4-1} % for checking your page length
\usepackage{graphicx}
\usepackage{epstopdf}
\usepackage{color}
\usepackage{amsmath}
\usepackage{booktabs}

\begin{document}
%\preprint{}

\title{Active demultiplexing of single-photons from a solid-state source} %Title of paper


\author{Francesco Lenzini}
%\email[]{}
%\thanks{}
\affiliation{Centre for Quantum Dynamics, Griffith University, Brisbane, Australia}

\author{Ben Haylock}
%\email[]{benjamin.haylock2@griffithuni.edu.au}
%\thanks{}
\affiliation{Centre for Quantum Dynamics, Griffith University, Brisbane, Australia}

\author{Juan C. Loredo}
%\email[]{}
%\thanks{}
\affiliation{Centre for Engineered Quantum Systems, Centre for Quantum Computer and Communication Technology, School of Mathematics and Physics, University of Queensland, Brisbane, Queensland 4072, Australia}
\affiliation{CNRS-C2N Centre de Nanosciences et de Nanotechnologies, Universit\'e Paris-Saclay, 91460 Marcoussis, France}

\author{Raphael A. Abrah\~ao}
%\email[]{}
%\thanks{}
\affiliation{Centre for Engineered Quantum Systems, Centre for Quantum Computer and Communication Technology, School of Mathematics and Physics, University of Queensland, Brisbane, Queensland 4072, Australia}

\author{Nor A. Zakaria}
%\email[]{}
%\thanks{}
\affiliation{Centre for Engineered Quantum Systems, Centre for Quantum Computer and Communication Technology, School of Mathematics and Physics, University of Queensland, Brisbane, Queensland 4072, Australia}

\author{Sachin Kasture}
%\email[]{benjamin.haylock2@griffithuni.edu.au}
%\thanks{}
\affiliation{Centre for Quantum Dynamics, Griffith University, Brisbane, Australia}

\author{Isabelle Sagnes}
%\email[]{}
%\thanks{}
\affiliation{CNRS-C2N Centre de Nanosciences et de Nanotechnologies, Universit\'e Paris-Saclay, 91460 Marcoussis, France}

\author{Aristide Lemaitre}
%\email[]{}
%\thanks{}
\affiliation{CNRS-C2N Centre de Nanosciences et de Nanotechnologies, Universit\'e Paris-Saclay, 91460 Marcoussis, France}

\author{Hoang-Phuong Phan}
\affiliation{Queensland Micro and Nanotechnology Centre, Griffith University, Brisbane, Australia}
\author{Dzung Viet Dao}
\affiliation{Queensland Micro and Nanotechnology Centre, Griffith University, Brisbane, Australia}
\affiliation{School of Engineering, Griffith University, Queensland, Australia}

\author{Pascale Senellart}
%\email[]{}
%\thanks{}
\affiliation{CNRS-C2N Centre de Nanosciences et de Nanotechnologies, Universit\'e Paris-Saclay, 91460 Marcoussis, France}
\affiliation{D\'epartement de Physique, Ecole Polytechnique, Universit\'e Paris-Saclay, F-91128 Palaiseau, France}

\author{Marcelo P. Almeida}
%\email[]{}
%\thanks{}
\affiliation{Centre for Engineered Quantum Systems, Centre for Quantum Computer and Communication Technology, School of Mathematics and Physics, University of Queensland, Brisbane, Queensland 4072, Australia}

\author{Andrew G. White}
%\email[]{}
%\thanks{}
\affiliation{Centre for Engineered Quantum Systems, Centre for Quantum Computer and Communication Technology, School of Mathematics and Physics, University of Queensland, Brisbane, Queensland 4072, Australia}

\author{Mirko Lobino}
%\email[]{}
%\thanks{}
\affiliation{Centre for Quantum Dynamics, Griffith University, Brisbane, Australia}
\affiliation{Queensland Micro and Nanotechnology Centre, Griffith University, Brisbane, Australia}

%\date{\today}

\begin{abstract}
A scheme for active temporal-to-spatial demultiplexing of single-photons generated by a solid-state source is introduced. The scheme scales quasi-polynomially with photon number, providing a viable technological path for routing $n$ photons in the one temporal stream from a single emitter to $n$ different spatial modes. The active demultiplexing is demonstrated using a state-of-the-art photon source---a quantum-dot deterministically coupled to a micropillar cavity---and a custom-built demultiplexer---a network of electro-optically reconfigurable waveguides monolithically integrated in a lithium niobate chip. The measured demultiplexer performance can enable a six-photon rate three orders of magnitude higher than the equivalent heralded SPDC source, providing a platform for intermediate quantum computation protocols.
\end{abstract}

%\pacs{07,42}% insert suggested PACS numbers in braces on next line

\maketitle %\maketitle must follow title, authors, abstract and \pacs

A key requirement for large-scale quantum photonic technologies is the availability of reliable sources of multiple indistinguishable single-photons.
To date, spontaneous parametric down-conversion (SPDC) sources have been the most widely-used technology in the generation of indistinguishable single-photons.
However, the presence of unwanted multiple-photon
terms in the SPDC state limits the brightness of high-purity single-photon sources to values lower than $1 \%$~\cite{Weinhold:09}. To circumvent this limitation different approaches have been introduced, including active spatial~\cite{Ma:11, Collins:13}, temporal~\cite{Kaneda:15, Xiong:16}, and spatio-temporal~\cite{Tanzilli:14, Mendoza:16} multiplexing schemes that combine the outputs of many SPDC sources to create one bright source without deteriorating single-photon purities---although typically at the cost of a large resource overhead.

In comparison, single emitters have the advantage of producing nearly-pure single-photon Fock states. %~\cite{Eisaman:11}. 
Very recent advances in quantum dot (QD) technologies have resulted in single-photon sources with simultaneously near-perfect purity, near-unity indistinguishability, and high efficiencies~\cite{Somaschi:16,Ding:16}---over an order of magnitude brighter than SPDC sources with equivalent levels of purity and indistinguishability. Thus, quantum dots have now become an attractive platform to develop multi-fold single-photon (multi-photon) sources.

Achieving high indistinguishability and brightness with multiple independent QDs is still a challenge. However, it has been shown that a single QD coupled to a micropillar cavity can emit photons with excellent indistinguishability over long emission timescales~\cite{Loredo:16, Wang:16}, meaning that temporal-to-spatial demultiplexing can be used to obtain 
multi-photon sources. In this work, we implement two important advances towards the realisation of a scalable multi-fold single-photon source. We first demonstrate the active temporal-to-spatial demultiplexing of a stream of photons 
to create multi-photon sources with small resource overhead. Secondly, we introduce the first integrated zero buffer active spatial and temporal photonic demultiplexing device, suitable for use with any high brightness source.

\begin{figure}[htbp]
	\centering
	\includegraphics[width=\linewidth]{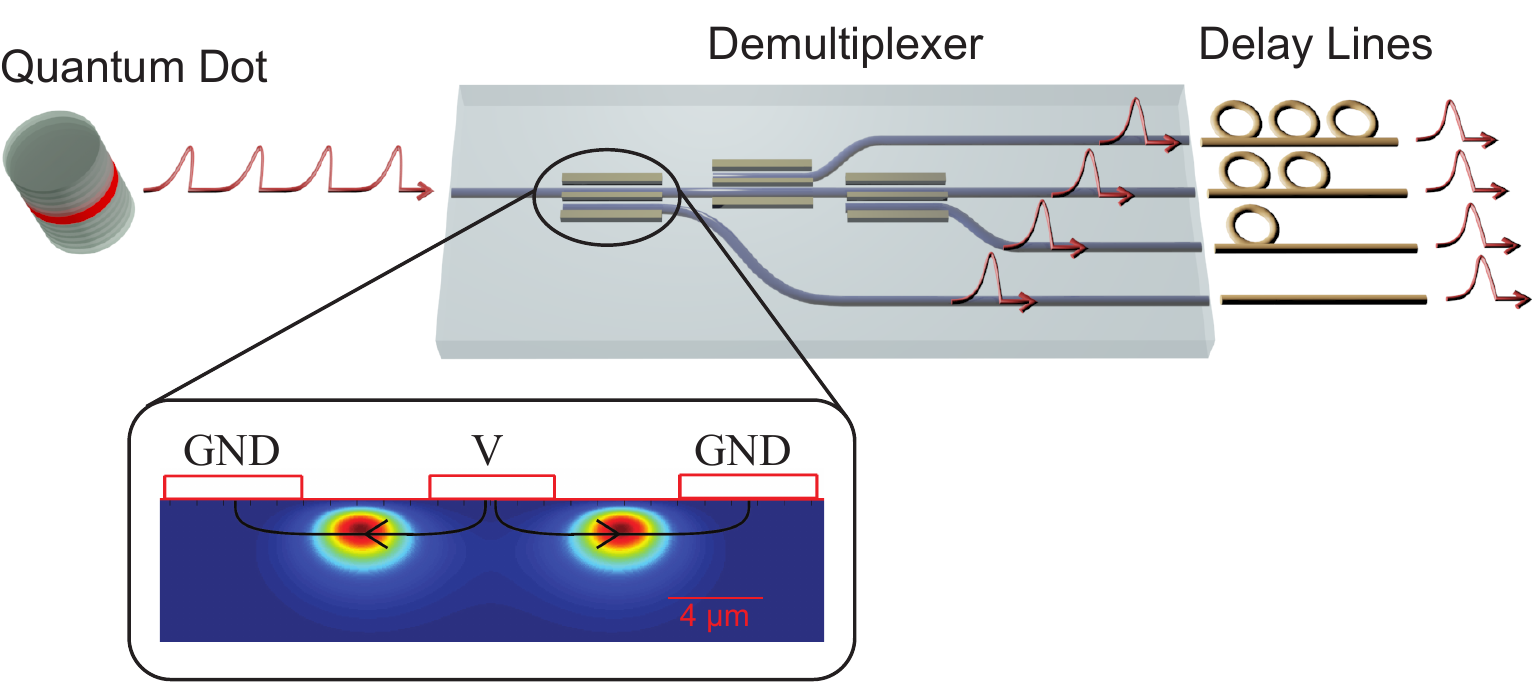}
	\caption{Scheme for active spatial-temporal demultiplexing of single-photons generated by a solid-state source. A stream of single-photons emitted at successive time intervals from a single emitter (here a quantum dot couple to a micropillar cavity) are actively routed into different spatial channels by an optical demultiplexer. A set of delay lines at the output can be used to match the different arrival times of the emitted single photons. The optical demultiplexer consists of a network of reconfigurable directional coupler waveguides with electro-optically tunable splitting ratio. The inset shows the configuration of the electrodes in each directional coupler. The colormap (a.u.) represents the intensity mode profiles at 932 nm in the waveguide structure calculated with the theoretical model from Ref.~\citenum{Lenzini:16} and the black arrows show the direction of the applied electric field.}
	\label{fig:1}
\end{figure}
Figure~\ref{fig:1} schematically depicts our proposed demultiplexing protocol. A temporal stream of single-photons emitted from a quantum dot-micropillar system is actively routed into different spatial channels by an optical demultiplexer. The demultiplexer is an integrated waveguide device with one input and four output channels made of a network of electro-optically reconfigurable directional couplers fabricated on an X-cut lithium niobate substrate by the annealed proton exchange technique~\cite{Lenzini:16}. 
Electrodes are patterned on top of the waveguides as shown in the inset of Fig.~\ref{fig:1}, and can be used to tune the splitting ratio in the full $0-100\%$ range by changing the phase mismatch $\Delta \beta$ between interacting modes~\cite{Taylor:17}. Monolithic integration of the directional coupler network on a single chip is necessary for reduced insertion losses, and with our technology it allows up to 10 output channels in a 5 cm long device. 

The $n$-photon count rate $c_\text{DM}(n)$ measured at the output of an n-channel demultiplexer can be expressed as 
\begin{equation}
c_\text{DM}(n)=R\left[\eta_\text{SD}\eta_\text{det}\right]^n S_\text{DM}(n),
\label{Eq:rate}
\end{equation}
where $\eta_\text{SD}= \eta_{\text{QD}} T$ is the product of the 
source brightness {$\eta_{\text{QD}}$} at the input of the demultiplexer times the total transmission of the device $T$, $R$ is the pump rate of the source and $\eta_\text{det}$ is the efficiency of the detectors. $S_\text{DM}(n)$ is a parameter which accounts for how the efficiency of the demultiplexing scheme scales with increasing number of photons. Note that the term $\left[\eta_\text{SD}\eta_\text{det}\right]^n$ is intrinsically probabilistic, and will unavoidably result in an exponential decay with photon number. In a probabilistic scheme~\cite{Loredo:16Boson} --made of a network of passive beam splitters-- the demultiplexing parameter scales as $S_\text{DM}(n){=} (1{/}n)^{n}$, super-exponentially decreasing with $n$---a non-scalable approach. In contrast, in an active demultiplexing scheme the scaling is 

\begin{equation}
S_\text{DM}{(n)}{=} \frac{1}{n} \left[\eta_{\text{DM}}^n + (n-1)\left(\frac{1-\eta_{\text{DM}}}{n-1}\right)^n \right] ,
\end{equation}
where $\eta_{\text{DM}}$ is the ``switching efficiency", defined as the average probability of routing a single photon in the desired channel in each time bin. In the limit of \textit{deterministic} demultiplexing, i.e. $\eta_{\text{DM}}{\to}1$, the scaling becomes polynomial in $n$---thus constituting a \textit{scalable} approach. 

The waveguides were fabricated with a 6~$\mathrm{\mu m}$ channel width and a proton exchange depth of 0.47~$\mathrm{\mu m}$ followed by annealing in air at 328~$^\circ$C for 15~h. These parameters are chosen in order to ensure good overlap with single-mode fiber and single-mode operation at {$\sim$}930~nm, the emission wavelength of our InGaAs QD. Each directional coupler has a distance between waveguide centres of 8.8~$\mathrm{\mu m}$ and a 4.5 mm length (equal to three coupling lengths), resulting in complete transmission of light into the coupled waveguide when no voltage to the corresponding switching electrodes is applied.

\begin{figure*}[htbp]
	\centering
	\includegraphics[width=\linewidth]{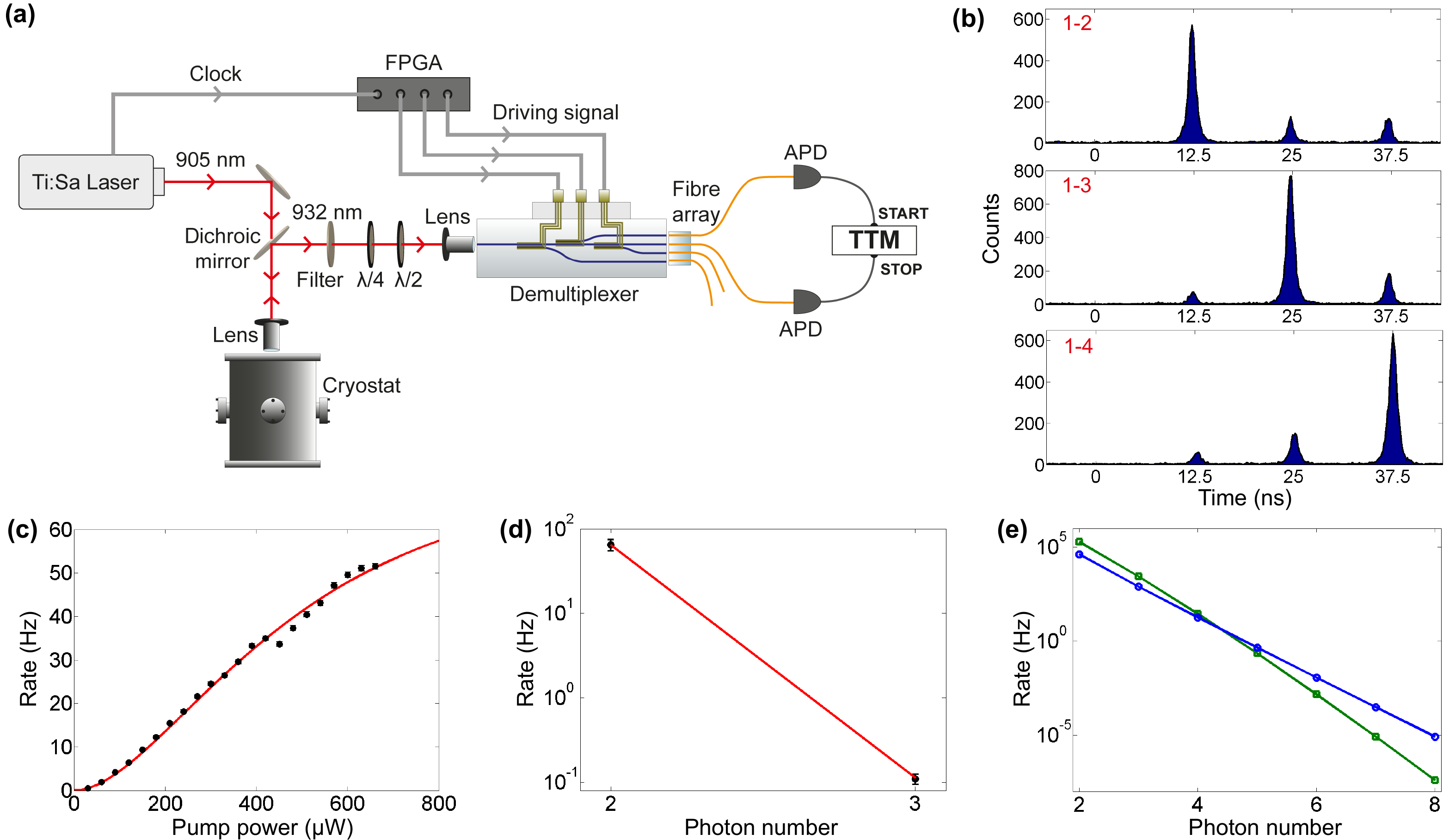}
	\caption{(a): Setup for the experimental implementation of the demultiplexing scheme (detailed description is given in the main text). (b): Time histograms for the two-photon coincidences between the first output and all other channels for a pump power $P=660 \ \mu \mbox{W}$ and a 2 minute acquisition time (waveguide numbering is from top to bottom). (c): Measured two-photon coincidence rates $c_p(2)$ as a function of the pump power $P$. Red line is the fit made with the saturation function given in the main text. Error bars are smaller than data points. (d): Measured two-photon and three-photon coincidence rates for a pump power $P=660 ~\mu \mbox{W}$. Red line is the fit made with the function in Eq.~\ref{Eq:rate}. (e): Comparison between the estimated photon rates at the output of the demultiplexer of an active (blue points) and probabilistic (green points) demultiplexing schemes for a state-of-the art QD pumped under resonant excitation~\cite{Somaschi:16}.}
	\label{fig:2}
\end{figure*}

The performance of the demultiplexer is tested in conjunction with a single-photon source based on a QD deterministically coupled to a micropillar cavity~\cite{Gazzano:2013aa,Loredo:16}. The experimental setup is schematically shown in Fig.~\ref{fig:2}a: the QD is quasi-resonantly pumped via p-shell excitation with a 905 nm 80 MHz 5 ps pulsed Ti:Sapph laser. The single-photons have a 932~nm emission wavelength and are separated from the pump beam via a dichroic mirror and a $0.85$~nm FWHM bandpass filter. Quarter- and half-wave plates are used at the input for polarisation alignment as the waveguides within the demultiplexer guide one (horizontal) polarisation. 
In our case, this reduces the available photon flux at the input of the demultiplexer by ${\sim}50\%$ since the source is only weakly polarised~\cite{Gazzano:2013aa}, an issue absent if operated with sources engineered to exhibit a large degree of polarization.
Photons are injected at the input of the device with a lens of NA=0.55 and all four outputs are collected with a fibre V-Groove array. Photon-coincidences between the output channels are measured using avalanche photodiodes with $30\%$ average quantum efficiency, and a time-tagging module (TTM). 
The electrodes of the demultiplexer are driven with a custom-made pulse generator based on a field programmable gate array (FPGA)~\cite{Ben:23}, which produces a temporal sequence of rectangular pulses with varying amplitude voltage synchronized with the pulsing of the Ti:Sapph laser.

To verify the correct operation of the switches, as well as their synchronization with the master laser, we first reconstruct the time histograms of two-photon coincidences counts between the first output of the demultiplexer and all other channels. The device is cyclically operated such that the first photon is sent to output one, the second to output two, and so on, and coincidences are measured between all four outputs simultaneously. Figure~\ref{fig:2}b shows the three time histograms (from a total of six pair wise combinations) of the coincidences measured by all four detectors. We observe enhanced peaks in coincidences at the corresponding delays of our demultiplexer, together with suppressed counts at different delays---showing the correct functioning of our device. 
The non-vanishing coincidence counts (smaller peaks) in the histograms arise from imperfect device operation, due to the performance of the modulated couplers. From the data in Fig.~\ref{fig:2}b we calculated the splitting ratios of the three switches for both settings (see Tab.~\ref{table_splitting}). The absence of counts at zero time delay (at the same level of accidental counts) is due to the low $g^2(0)$ value of the source, measured as $g^{2}(0){=}0.029{\pm}0.001$ at $P{=}3P_0$ in \cite{Loredo:16}.

\begin{table}[b]
	\begin{center}
		\begin{tabular}{c| c c c}
			\hline
			switch & 1 & 2 & 3 \\
			\hline
			$on$ &0.87$\pm$0.06 &0.94$\pm$0.05 &0.90$\pm$0.06\\
			$off$ &0.06$\pm$0.02 &0.13$\pm$0.03 &0.13$\pm$0.05\\
			\hline
		\end{tabular}
		\vspace{0.25cm} \caption{\footnotesize\label{table_splitting}
			Splitting ratios of the directional couplers calculated from the data in Fig.\ref{fig:2}b., with uncertainty from the fit confidence. Non-zero $off$ values are caused by incorrect driving voltages, and non-unity $on$ values by waveguide imperfections.}
	\end{center}
\end{table}

Figure.~\ref{fig:2}c shows the measured power-dependent rate of two-photon coincidences $c_\text{DM}(2)$ at outputs 1 and 2 of our demultiplexer. As expected for a QD pumped under quasi-resonant excitation it follows a saturation function $c_\text{DM}(2){=}c_\text{max}(2)\left[1{-}\text{exp}(-P/P_0)\right]^2$ quadratic in the $P$-dependance of the single-photon brightness. A fit to the data results in $c_\text{max}(2){=}70.9{\pm}3.0$~Hz, the maximum detected 2-photon rate, and $P_0{=}348{\pm}16$ $\mu$W the saturation power.
The switching efficiency $\eta_{\text{DM}}$ is finally estimated by measuring two-fold and three-fold photon coincidences at the output for a pump power $P{=}660 \ \mu \text{W}$ (Fig.~\ref{fig:2}d) and by fitting the experimental points with Eq.~\ref{Eq:rate}, where $R=80 \ \mbox{MHz}$, $\eta_{\text{det}}=30 \%$, and $\eta_{\text{SD}}=0.76 \%$ is calculated from the total number of counts measured with the four detectors. We find an average switching efficiency $\eta_{\text{DM}}=0.78\pm 0.06$, in good agreement with the value $\eta_{\text{DM}}=0.80 \pm 0.09$ predicted from the measured splitting ratios. Four-fold coincidences were deteriorated due to the low value of $\eta_{\text{SD}}$ in the current version of our system, not producing sufficient statistics in the given acquisition time.

To investigate the potential of our technology for the realisation of a multi-photon source with larger numbers we calculate the expected photon rates at the output of the demultiplexer for a state-of-the-art QD with $15\%$ polarised brightness pumped under resonant-excitation \cite{Somaschi:16}. The total transmission of our demultiplexer is tested by coupling the waveguide with a gaussian mode from an single-mode optical fibre and is found to be $T{=}30 \%$. This value is compatible with an overlap with the waveguide mode $\simeq 85 \%$, as measured from mode imaging at the output of the waveguide, $14 \%$ Fresnel losses at the input and output facet and propagation losses $\simeq 0.65 \ \mbox{dB/cm}$. In Fig.~\ref{fig:2}e we report the expected photon rates for increasing photon numbers calculated for a pump rate $R=80 \ \mbox{MHz}$, $\eta_{\text{DM}}=78\%$, and a transmission $T = 0.3/(0.86 \times 0.86)$ corrected for Fresnel losses, that can be eliminated with an anti-reflection coating at the input and output facets. The QD brightness is corrected by an additional loss factor $65\%$ that takes into account the coupling efficiency of the QD emission mode to a single-mode fibre \cite{Loredo:16}. 
The proposed system with these parameters is expected to outperform a probabilistic demultiplexing scheme---made of a network of passive beam splitters with zero propagation losses-- for a number of photons $n{>}4$ and would enable a 6-photon rate $\simeq 0.01 \ \mbox{Hz}$, which is three orders of magnitude larger than what could be obtained with six heralded SPDC sources with equivalent quality~\cite{Somaschi:16}. The same calculation for a resonantly-excited QD with $14\%$ brightness measured at the ouptut of a single-mode fibre~\cite{Ding:16}, would enable, instead, a 6-photon rate $\simeq 0.1 \ \mbox{Hz}$. This technology offers great potential for further improvement, in particular by the use of the Reverse Proton exchange technique~\cite{Lenzini:16} for an improved coupling with optical fibres and reduced surface-scattering losses we estimate that we can achieve insertion losses lower than 3 dB. Such upgrades will enable the scaling of this platform to a larger number of photons.

In conclusion, we have proposed and experimentally implemented the first example of active demultiplexing with a single integrated device of single-photons from a solid-state source. The performance of the demultiplexer has been analysed in conjunction with a QD pumped under quasi-resonant excitation and we have discussed the potential of our technology for state-of-the art quantum dots. The proposed demultiplexing device is of general interest for any bright temporally distributed single-photon source and provides a scalable approach for the realisation of multi-photon sources of larger photon numbers. Our platform thus constitutes a very promising approach for scalable quantum photonics, in particular for protocols of intermediate---i.e., non universal---quantum computation, such as Boson Sampling~\cite{AAronson:18, AAronson:19, Shen:20, Lund:21, Huh:22}, where, arguably, as few as seven photons from an actively demultiplexed quantum dot-based source could finally demonstrate the quantum advantage over classical systems~\cite{QSup:Latmiral16}.

\small
\textbf{Acknowledgements.} This work has been supported by the Australian Research Council (ARC) under the Grant No. DP140100808 and  fellowship DE120101899, the Centres of Excellence for Engineered Quantum Systems (EQuS) and Quantum Computing and Communication Technology (CQC2T), and the Griffith University Research Infrastructure Program. %%AGW Added COE acknowledgements.
This work was partially supported by the ERC Starting Grant No. 277885 QD-CQED, the French Agence Nationale pour la Recherche (grant ANR DELIGHT), the French RENATECH network and a public grant overseen by the French National Research Agency (ANR) as part of the ”Investissements d’Avenir” program (Labex NanoSaclay, reference: ANR-10-LABX- 0035). %%Pascale aknowledgements
This work was performed in part at the Queensland node of the Australian National Fabrication Facility, a company established under the National Collaborative Research Infrastructure Strategy to provide nano- and microfabrication facilities for Australia's %%AGW Cut "Australiaâs"
researchers.

% Bibliography
\bibliography{Lenzini_bibliography}

\begin{thebibliography}{22}%
\makeatletter
\providecommand \@ifxundefined [1]{%
 \@ifx{#1\undefined}
}%
\providecommand \@ifnum [1]{%
 \ifnum #1\expandafter \@firstoftwo
 \else \expandafter \@secondoftwo
 \fi
}%
\providecommand \@ifx [1]{%
 \ifx #1\expandafter \@firstoftwo
 \else \expandafter \@secondoftwo
 \fi
}%
\providecommand \natexlab [1]{#1}%
\providecommand \enquote  [1]{``#1''}%
\providecommand \bibnamefont  [1]{#1}%
\providecommand \bibfnamefont [1]{#1}%
\providecommand \citenamefont [1]{#1}%
\providecommand \href@noop [0]{\@secondoftwo}%
\providecommand \href [0]{\begingroup \@sanitize@url \@href}%
\providecommand \@href[1]{\@@startlink{#1}\@@href}%
\providecommand \@@href[1]{\endgroup#1\@@endlink}%
\providecommand \@sanitize@url [0]{\catcode `\\12\catcode `\$12\catcode
  `\&12\catcode `\#12\catcode `\^12\catcode `\_12\catcode `\%12\relax}%
\providecommand \@@startlink[1]{}%
\providecommand \@@endlink[0]{}%
\providecommand \url  [0]{\begingroup\@sanitize@url \@url }%
\providecommand \@url [1]{\endgroup\@href {#1}{\urlprefix }}%
\providecommand \urlprefix  [0]{URL }%
\providecommand \Eprint [0]{\href }%
\providecommand \doibase [0]{http://dx.doi.org/}%
\providecommand \selectlanguage [0]{\@gobble}%
\providecommand \bibinfo  [0]{\@secondoftwo}%
\providecommand \bibfield  [0]{\@secondoftwo}%
\providecommand \translation [1]{[#1]}%
\providecommand \BibitemOpen [0]{}%
\providecommand \bibitemStop [0]{}%
\providecommand \bibitemNoStop [0]{.\EOS\space}%
\providecommand \EOS [0]{\spacefactor3000\relax}%
\providecommand \BibitemShut  [1]{\csname bibitem#1\endcsname}%
\let\auto@bib@innerbib\@empty
%</preamble>
\bibitem [{\citenamefont {Barbieri}\ \emph {et~al.}(2009)\citenamefont
  {Barbieri}, \citenamefont {Weinhold}, \citenamefont {Lanyon}, \citenamefont
  {Gilchrist}, \citenamefont {Resch}, \citenamefont {Almeida},\ and\
  \citenamefont {White}}]{Weinhold:09}%
  \BibitemOpen
  \bibfield  {author} {\bibinfo {author} {\bibfnamefont {M.}~\bibnamefont
  {Barbieri}}, \bibinfo {author} {\bibfnamefont {T.}~\bibnamefont {Weinhold}},
  \bibinfo {author} {\bibfnamefont {B.}~\bibnamefont {Lanyon}}, \bibinfo
  {author} {\bibfnamefont {A.}~\bibnamefont {Gilchrist}}, \bibinfo {author}
  {\bibfnamefont {K.}~\bibnamefont {Resch}}, \bibinfo {author} {\bibfnamefont
  {M.}~\bibnamefont {Almeida}}, \ and\ \bibinfo {author} {\bibfnamefont
  {A.}~\bibnamefont {White}},\ }\href {\doibase 10.1080/09500340802337374}
  {\bibfield  {journal} {\bibinfo  {journal} {Journal of Modern Optics}\
  }\textbf {\bibinfo {volume} {56}},\ \bibinfo {pages} {209} (\bibinfo {year}
  {2009})},\ \Eprint
  {http://arxiv.org/abs/http://dx.doi.org/10.1080/09500340802337374}
  {http://dx.doi.org/10.1080/09500340802337374} \BibitemShut {NoStop}%
\bibitem [{\citenamefont {Ma}\ \emph {et~al.}(2011)\citenamefont {Ma},
  \citenamefont {Zotter}, \citenamefont {Kofler}, \citenamefont {Jennewein},\
  and\ \citenamefont {Zeilinger}}]{Ma:11}%
  \BibitemOpen
  \bibfield  {author} {\bibinfo {author} {\bibfnamefont {X.-s.}\ \bibnamefont
  {Ma}}, \bibinfo {author} {\bibfnamefont {S.}~\bibnamefont {Zotter}}, \bibinfo
  {author} {\bibfnamefont {J.}~\bibnamefont {Kofler}}, \bibinfo {author}
  {\bibfnamefont {T.}~\bibnamefont {Jennewein}}, \ and\ \bibinfo {author}
  {\bibfnamefont {A.}~\bibnamefont {Zeilinger}},\ }\href {\doibase
  10.1103/PhysRevA.83.043814} {\bibfield  {journal} {\bibinfo  {journal} {Phys.
  Rev. A}\ }\textbf {\bibinfo {volume} {83}},\ \bibinfo {pages} {043814}
  (\bibinfo {year} {2011})}\BibitemShut {NoStop}%
\bibitem [{\citenamefont {Collins}\ \emph {et~al.}(2013)\citenamefont
  {Collins}, \citenamefont {Xiong}, \citenamefont {Rey}, \citenamefont {Vo},
  \citenamefont {He}, \citenamefont {Shahnia}, \citenamefont {Reardon},
  \citenamefont {Krauss}, \citenamefont {Steel}, \citenamefont {Clark},\ and\
  \citenamefont {Eggleton}}]{Collins:13}%
  \BibitemOpen
  \bibfield  {author} {\bibinfo {author} {\bibfnamefont {M.}~\bibnamefont
  {Collins}}, \bibinfo {author} {\bibfnamefont {C.}~\bibnamefont {Xiong}},
  \bibinfo {author} {\bibfnamefont {I.}~\bibnamefont {Rey}}, \bibinfo {author}
  {\bibfnamefont {T.}~\bibnamefont {Vo}}, \bibinfo {author} {\bibfnamefont
  {J.}~\bibnamefont {He}}, \bibinfo {author} {\bibfnamefont {S.}~\bibnamefont
  {Shahnia}}, \bibinfo {author} {\bibfnamefont {C.}~\bibnamefont {Reardon}},
  \bibinfo {author} {\bibfnamefont {T.}~\bibnamefont {Krauss}}, \bibinfo
  {author} {\bibfnamefont {M.}~\bibnamefont {Steel}}, \bibinfo {author}
  {\bibfnamefont {A.}~\bibnamefont {Clark}}, \ and\ \bibinfo {author}
  {\bibfnamefont {B.}~\bibnamefont {Eggleton}},\ }\href@noop {} {\bibfield
  {journal} {\bibinfo  {journal} {Nat. Comms.}\ }\textbf {\bibinfo {volume}
  {4}} (\bibinfo {year} {2013})}\BibitemShut {NoStop}%
\bibitem [{\citenamefont {Kaneda}\ \emph {et~al.}(2015)\citenamefont {Kaneda},
  \citenamefont {Christensen}, \citenamefont {Wong}, \citenamefont {Park},
  \citenamefont {McCusker},\ and\ \citenamefont {Kwiat}}]{Kaneda:15}%
  \BibitemOpen
  \bibfield  {author} {\bibinfo {author} {\bibfnamefont {F.}~\bibnamefont
  {Kaneda}}, \bibinfo {author} {\bibfnamefont {B.~G.}\ \bibnamefont
  {Christensen}}, \bibinfo {author} {\bibfnamefont {J.~J.}\ \bibnamefont
  {Wong}}, \bibinfo {author} {\bibfnamefont {H.~S.}\ \bibnamefont {Park}},
  \bibinfo {author} {\bibfnamefont {K.~T.}\ \bibnamefont {McCusker}}, \ and\
  \bibinfo {author} {\bibfnamefont {P.~G.}\ \bibnamefont {Kwiat}},\ }\href
  {\doibase 10.1364/OPTICA.2.001010} {\bibfield  {journal} {\bibinfo  {journal}
  {Optica}\ }\textbf {\bibinfo {volume} {2}},\ \bibinfo {pages} {1010}
  (\bibinfo {year} {2015})}\BibitemShut {NoStop}%
\bibitem [{\citenamefont {Xiong}\ \emph {et~al.}(2013)\citenamefont {Xiong},
  \citenamefont {Zhang}, \citenamefont {Liu}, \citenamefont {Collins},
  \citenamefont {Mahendra}, \citenamefont {Helt}, \citenamefont {Steel},
  \citenamefont {Choi}, \citenamefont {Chae}, \citenamefont {Leong},\ and\
  \citenamefont {Eggleton}}]{Xiong:16}%
  \BibitemOpen
  \bibfield  {author} {\bibinfo {author} {\bibfnamefont {C.}~\bibnamefont
  {Xiong}}, \bibinfo {author} {\bibfnamefont {X.}~\bibnamefont {Zhang}},
  \bibinfo {author} {\bibfnamefont {Z.}~\bibnamefont {Liu}}, \bibinfo {author}
  {\bibfnamefont {M.~J.}\ \bibnamefont {Collins}}, \bibinfo {author}
  {\bibfnamefont {A.}~\bibnamefont {Mahendra}}, \bibinfo {author}
  {\bibfnamefont {L.~G.}\ \bibnamefont {Helt}}, \bibinfo {author}
  {\bibfnamefont {M.~J.}\ \bibnamefont {Steel}}, \bibinfo {author}
  {\bibfnamefont {D.~Y.}\ \bibnamefont {Choi}}, \bibinfo {author}
  {\bibfnamefont {C.~J.}\ \bibnamefont {Chae}}, \bibinfo {author}
  {\bibfnamefont {P.~H.~W.}\ \bibnamefont {Leong}}, \ and\ \bibinfo {author}
  {\bibfnamefont {B.~J.}\ \bibnamefont {Eggleton}},\ }\href@noop {} {\bibfield
  {journal} {\bibinfo  {journal} {Nat. Comms.}\ }\textbf {\bibinfo {volume}
  {7}} (\bibinfo {year} {2013})}\BibitemShut {NoStop}%
\bibitem [{\citenamefont {Meany}\ \emph {et~al.}(2014)\citenamefont {Meany},
  \citenamefont {Ngah}, \citenamefont {Collins}, \citenamefont {Clark},
  \citenamefont {Williams}, \citenamefont {Eggleton}, \citenamefont {Steel},
  \citenamefont {Withford}, \citenamefont {Alibart},\ and\ \citenamefont
  {Tanzilli}}]{Tanzilli:14}%
  \BibitemOpen
  \bibfield  {author} {\bibinfo {author} {\bibfnamefont {T.}~\bibnamefont
  {Meany}}, \bibinfo {author} {\bibfnamefont {L.~A.}\ \bibnamefont {Ngah}},
  \bibinfo {author} {\bibfnamefont {M.~J.}\ \bibnamefont {Collins}}, \bibinfo
  {author} {\bibfnamefont {A.~S.}\ \bibnamefont {Clark}}, \bibinfo {author}
  {\bibfnamefont {R.~J.}\ \bibnamefont {Williams}}, \bibinfo {author}
  {\bibfnamefont {B.~J.}\ \bibnamefont {Eggleton}}, \bibinfo {author}
  {\bibfnamefont {M.~J.}\ \bibnamefont {Steel}}, \bibinfo {author}
  {\bibfnamefont {M.~J.}\ \bibnamefont {Withford}}, \bibinfo {author}
  {\bibfnamefont {O.}~\bibnamefont {Alibart}}, \ and\ \bibinfo {author}
  {\bibfnamefont {S.}~\bibnamefont {Tanzilli}},\ }\href@noop {} {\bibfield
  {journal} {\bibinfo  {journal} {Laser Photon. Rev.}\ }\textbf {\bibinfo
  {volume} {8}},\ \bibinfo {pages} {L42} (\bibinfo {year} {2014})}\BibitemShut
  {NoStop}%
\bibitem [{\citenamefont {Mendoza}\ \emph {et~al.}(2016)\citenamefont
  {Mendoza}, \citenamefont {Santagati}, \citenamefont {Munns}, \citenamefont
  {Hemsley}, \citenamefont {Piekarek}, \citenamefont {Mart\'{i}n-L\'{o}pez},
  \citenamefont {Marshall}, \citenamefont {Bonneau}, \citenamefont {Thompson},\
  and\ \citenamefont {O'Brien}}]{Mendoza:16}%
  \BibitemOpen
  \bibfield  {author} {\bibinfo {author} {\bibfnamefont {G.~J.}\ \bibnamefont
  {Mendoza}}, \bibinfo {author} {\bibfnamefont {R.}~\bibnamefont {Santagati}},
  \bibinfo {author} {\bibfnamefont {J.}~\bibnamefont {Munns}}, \bibinfo
  {author} {\bibfnamefont {E.}~\bibnamefont {Hemsley}}, \bibinfo {author}
  {\bibfnamefont {M.}~\bibnamefont {Piekarek}}, \bibinfo {author}
  {\bibfnamefont {E.}~\bibnamefont {Mart\'{i}n-L\'{o}pez}}, \bibinfo {author}
  {\bibfnamefont {G.~D.}\ \bibnamefont {Marshall}}, \bibinfo {author}
  {\bibfnamefont {D.}~\bibnamefont {Bonneau}}, \bibinfo {author} {\bibfnamefont
  {M.~G.}\ \bibnamefont {Thompson}}, \ and\ \bibinfo {author} {\bibfnamefont
  {J.~L.}\ \bibnamefont {O'Brien}},\ }\href {\doibase 10.1364/OPTICA.3.000127}
  {\bibfield  {journal} {\bibinfo  {journal} {Optica}\ }\textbf {\bibinfo
  {volume} {3}},\ \bibinfo {pages} {127} (\bibinfo {year} {2016})}\BibitemShut
  {NoStop}%
\bibitem [{\citenamefont {Somaschi}\ \emph {et~al.}(2016)\citenamefont
  {Somaschi}, \citenamefont {Giesz}, \citenamefont {Santis}, \citenamefont
  {Loredo}, \citenamefont {Almeida}, \citenamefont {Hornecker}, \citenamefont
  {Portalupi}, \citenamefont {Grange}, \citenamefont {Ant{\'o}n}, \citenamefont
  {Demory}, \citenamefont {G{\'o}mez}, \citenamefont {Sagnes}, \citenamefont
  {Lanzillotti-Kimura}, \citenamefont {Lema{\'\i}tre}, \citenamefont
  {Auffeves}, \citenamefont {White}, \citenamefont {Lanco},\ and\ \citenamefont
  {Senellart}}]{Somaschi:16}%
  \BibitemOpen
  \bibfield  {author} {\bibinfo {author} {\bibfnamefont {N.}~\bibnamefont
  {Somaschi}}, \bibinfo {author} {\bibfnamefont {V.}~\bibnamefont {Giesz}},
  \bibinfo {author} {\bibfnamefont {L.~D.}\ \bibnamefont {Santis}}, \bibinfo
  {author} {\bibfnamefont {J.~C.}\ \bibnamefont {Loredo}}, \bibinfo {author}
  {\bibfnamefont {M.~P.}\ \bibnamefont {Almeida}}, \bibinfo {author}
  {\bibfnamefont {G.}~\bibnamefont {Hornecker}}, \bibinfo {author}
  {\bibfnamefont {S.~L.}\ \bibnamefont {Portalupi}}, \bibinfo {author}
  {\bibfnamefont {T.}~\bibnamefont {Grange}}, \bibinfo {author} {\bibfnamefont
  {C.}~\bibnamefont {Ant{\'o}n}}, \bibinfo {author} {\bibfnamefont
  {J.}~\bibnamefont {Demory}}, \bibinfo {author} {\bibfnamefont
  {C.}~\bibnamefont {G{\'o}mez}}, \bibinfo {author} {\bibfnamefont
  {I.}~\bibnamefont {Sagnes}}, \bibinfo {author} {\bibfnamefont {N.~D.}\
  \bibnamefont {Lanzillotti-Kimura}}, \bibinfo {author} {\bibfnamefont
  {A.}~\bibnamefont {Lema{\'\i}tre}}, \bibinfo {author} {\bibfnamefont
  {A.}~\bibnamefont {Auffeves}}, \bibinfo {author} {\bibfnamefont {A.~G.}\
  \bibnamefont {White}}, \bibinfo {author} {\bibfnamefont {L.}~\bibnamefont
  {Lanco}}, \ and\ \bibinfo {author} {\bibfnamefont {P.}~\bibnamefont
  {Senellart}},\ }\href@noop {} {\bibfield  {journal} {\bibinfo  {journal}
  {Nat. Photon.}\ }\textbf {\bibinfo {volume} {10}},\ \bibinfo {pages} {340}
  (\bibinfo {year} {2016})}\BibitemShut {NoStop}%
\bibitem [{\citenamefont {Ding}\ \emph {et~al.}(2016)\citenamefont {Ding},
  \citenamefont {He}, \citenamefont {Duan}, \citenamefont {Gregersen},
  \citenamefont {Chen}, \citenamefont {Unsleber}, \citenamefont {Maier},
  \citenamefont {Schneider}, \citenamefont {Kamp}, \citenamefont {H\"ofling},
  \citenamefont {Lu},\ and\ \citenamefont {Pan}}]{Ding:16}%
  \BibitemOpen
  \bibfield  {author} {\bibinfo {author} {\bibfnamefont {X.}~\bibnamefont
  {Ding}}, \bibinfo {author} {\bibfnamefont {Y.}~\bibnamefont {He}}, \bibinfo
  {author} {\bibfnamefont {Z.-C.}\ \bibnamefont {Duan}}, \bibinfo {author}
  {\bibfnamefont {N.}~\bibnamefont {Gregersen}}, \bibinfo {author}
  {\bibfnamefont {M.-C.}\ \bibnamefont {Chen}}, \bibinfo {author}
  {\bibfnamefont {S.}~\bibnamefont {Unsleber}}, \bibinfo {author}
  {\bibfnamefont {S.}~\bibnamefont {Maier}}, \bibinfo {author} {\bibfnamefont
  {C.}~\bibnamefont {Schneider}}, \bibinfo {author} {\bibfnamefont
  {M.}~\bibnamefont {Kamp}}, \bibinfo {author} {\bibfnamefont {S.}~\bibnamefont
  {H\"ofling}}, \bibinfo {author} {\bibfnamefont {C.-Y.}\ \bibnamefont {Lu}}, \
  and\ \bibinfo {author} {\bibfnamefont {J.-W.}\ \bibnamefont {Pan}},\ }\href
  {\doibase 10.1103/PhysRevLett.116.020401} {\bibfield  {journal} {\bibinfo
  {journal} {Phys. Rev. Lett.}\ }\textbf {\bibinfo {volume} {116}},\ \bibinfo
  {pages} {020401} (\bibinfo {year} {2016})}\BibitemShut {NoStop}%
\bibitem [{\citenamefont {Loredo}\ \emph
  {et~al.}(2016{\natexlab{a}})\citenamefont {Loredo}, \citenamefont {Zakaria},
  \citenamefont {Somaschi}, \citenamefont {Anton}, \citenamefont {de~Santis},
  \citenamefont {Giesz}, \citenamefont {Grange}, \citenamefont {Broome},
  \citenamefont {Gazzano}, \citenamefont {Coppola}, \citenamefont {Sagnes},
  \citenamefont {Lemaitre}, \citenamefont {Auffeves}, \citenamefont
  {Senellart}, \citenamefont {Almeida},\ and\ \citenamefont
  {White}}]{Loredo:16}%
  \BibitemOpen
  \bibfield  {author} {\bibinfo {author} {\bibfnamefont {J.~C.}\ \bibnamefont
  {Loredo}}, \bibinfo {author} {\bibfnamefont {N.~A.}\ \bibnamefont {Zakaria}},
  \bibinfo {author} {\bibfnamefont {N.}~\bibnamefont {Somaschi}}, \bibinfo
  {author} {\bibfnamefont {C.}~\bibnamefont {Anton}}, \bibinfo {author}
  {\bibfnamefont {L.}~\bibnamefont {de~Santis}}, \bibinfo {author}
  {\bibfnamefont {V.}~\bibnamefont {Giesz}}, \bibinfo {author} {\bibfnamefont
  {T.}~\bibnamefont {Grange}}, \bibinfo {author} {\bibfnamefont {M.~A.}\
  \bibnamefont {Broome}}, \bibinfo {author} {\bibfnamefont {O.}~\bibnamefont
  {Gazzano}}, \bibinfo {author} {\bibfnamefont {G.}~\bibnamefont {Coppola}},
  \bibinfo {author} {\bibfnamefont {I.}~\bibnamefont {Sagnes}}, \bibinfo
  {author} {\bibfnamefont {A.}~\bibnamefont {Lemaitre}}, \bibinfo {author}
  {\bibfnamefont {A.}~\bibnamefont {Auffeves}}, \bibinfo {author}
  {\bibfnamefont {P.}~\bibnamefont {Senellart}}, \bibinfo {author}
  {\bibfnamefont {M.~P.}\ \bibnamefont {Almeida}}, \ and\ \bibinfo {author}
  {\bibfnamefont {A.~G.}\ \bibnamefont {White}},\ }\href {\doibase
  10.1364/OPTICA.3.000433} {\bibfield  {journal} {\bibinfo  {journal} {Optica}\
  }\textbf {\bibinfo {volume} {3}},\ \bibinfo {pages} {433} (\bibinfo {year}
  {2016}{\natexlab{a}})}\BibitemShut {NoStop}%
\bibitem [{\citenamefont {Wang}\ \emph {et~al.}(2016)\citenamefont {Wang},
  \citenamefont {Duan}, \citenamefont {Li}, \citenamefont {Chen}, \citenamefont
  {Li}, \citenamefont {He}, \citenamefont {Chen}, \citenamefont {He},
  \citenamefont {Ding}, \citenamefont {Peng}, \citenamefont {Schneider},
  \citenamefont {Kamp}, \citenamefont {H\"ofling}, \citenamefont {Lu},\ and\
  \citenamefont {Pan}}]{Wang:16}%
  \BibitemOpen
  \bibfield  {author} {\bibinfo {author} {\bibfnamefont {H.}~\bibnamefont
  {Wang}}, \bibinfo {author} {\bibfnamefont {Z.-C.}\ \bibnamefont {Duan}},
  \bibinfo {author} {\bibfnamefont {Y.-H.}\ \bibnamefont {Li}}, \bibinfo
  {author} {\bibfnamefont {S.}~\bibnamefont {Chen}}, \bibinfo {author}
  {\bibfnamefont {J.-P.}\ \bibnamefont {Li}}, \bibinfo {author} {\bibfnamefont
  {Y.-M.}\ \bibnamefont {He}}, \bibinfo {author} {\bibfnamefont {M.-C.}\
  \bibnamefont {Chen}}, \bibinfo {author} {\bibfnamefont {Y.}~\bibnamefont
  {He}}, \bibinfo {author} {\bibfnamefont {X.}~\bibnamefont {Ding}}, \bibinfo
  {author} {\bibfnamefont {C.-Z.}\ \bibnamefont {Peng}}, \bibinfo {author}
  {\bibfnamefont {C.}~\bibnamefont {Schneider}}, \bibinfo {author}
  {\bibfnamefont {M.}~\bibnamefont {Kamp}}, \bibinfo {author} {\bibfnamefont
  {S.}~\bibnamefont {H\"ofling}}, \bibinfo {author} {\bibfnamefont {C.-Y.}\
  \bibnamefont {Lu}}, \ and\ \bibinfo {author} {\bibfnamefont {J.-W.}\
  \bibnamefont {Pan}},\ }\href {\doibase 10.1103/PhysRevLett.116.213601}
  {\bibfield  {journal} {\bibinfo  {journal} {Phys. Rev. Lett.}\ }\textbf
  {\bibinfo {volume} {116}},\ \bibinfo {pages} {213601} (\bibinfo {year}
  {2016})}\BibitemShut {NoStop}%
\bibitem [{\citenamefont {Lenzini}\ \emph {et~al.}(2015)\citenamefont
  {Lenzini}, \citenamefont {Kasture}, \citenamefont {Haylock},\ and\
  \citenamefont {Lobino}}]{Lenzini:16}%
  \BibitemOpen
  \bibfield  {author} {\bibinfo {author} {\bibfnamefont {F.}~\bibnamefont
  {Lenzini}}, \bibinfo {author} {\bibfnamefont {S.}~\bibnamefont {Kasture}},
  \bibinfo {author} {\bibfnamefont {B.}~\bibnamefont {Haylock}}, \ and\
  \bibinfo {author} {\bibfnamefont {M.}~\bibnamefont {Lobino}},\ }\href@noop {}
  {\bibfield  {journal} {\bibinfo  {journal} {Opt. Express}\ }\textbf {\bibinfo
  {volume} {23}},\ \bibinfo {pages} {1748} (\bibinfo {year}
  {2015})}\BibitemShut {NoStop}%
\bibitem [{\citenamefont {Taylor}(1973)}]{Taylor:17}%
  \BibitemOpen
  \bibfield  {author} {\bibinfo {author} {\bibfnamefont {H.~F.}\ \bibnamefont
  {Taylor}},\ }\href@noop {} {\bibfield  {journal} {\bibinfo  {journal} {J.
  Appl. Phys.}\ }\textbf {\bibinfo {volume} {44}},\ \bibinfo {pages} {3257}
  (\bibinfo {year} {1973})}\BibitemShut {NoStop}%
\bibitem [{\citenamefont {Loredo}\ \emph
  {et~al.}(2016{\natexlab{b}})\citenamefont {Loredo}, \citenamefont {Broome},
  \citenamefont {Hilaire}, \citenamefont {Gazzano}, \citenamefont {Sagnes},
  \citenamefont {Lemaitre}, \citenamefont {Almeida}, \citenamefont
  {Senellart},\ and\ \citenamefont {White}}]{Loredo:16Boson}%
  \BibitemOpen
  \bibfield  {author} {\bibinfo {author} {\bibfnamefont {J.~C.}\ \bibnamefont
  {Loredo}}, \bibinfo {author} {\bibfnamefont {M.~A.}\ \bibnamefont {Broome}},
  \bibinfo {author} {\bibfnamefont {P.}~\bibnamefont {Hilaire}}, \bibinfo
  {author} {\bibfnamefont {O.}~\bibnamefont {Gazzano}}, \bibinfo {author}
  {\bibfnamefont {I.}~\bibnamefont {Sagnes}}, \bibinfo {author} {\bibfnamefont
  {A.}~\bibnamefont {Lemaitre}}, \bibinfo {author} {\bibfnamefont {M.~P.}\
  \bibnamefont {Almeida}}, \bibinfo {author} {\bibfnamefont {P.}~\bibnamefont
  {Senellart}}, \ and\ \bibinfo {author} {\bibfnamefont {A.~G.}\ \bibnamefont
  {White}},\ }\href@noop {} {\bibfield  {journal} {\bibinfo  {journal}
  {arXiv:1603.00054}\ } (\bibinfo {year} {2016}{\natexlab{b}})}\BibitemShut
  {NoStop}%
\bibitem [{\citenamefont {Gazzano}\ \emph {et~al.}(2013)\citenamefont
  {Gazzano}, \citenamefont {de~Vasconcellos}, \citenamefont {Arnold},
  \citenamefont {Nowak}, \citenamefont {Galopin}, \citenamefont {Sagnes},
  \citenamefont {Lanco}, \citenamefont {Lema{\^\i}tre},\ and\ \citenamefont
  {Senellart}}]{Gazzano:2013aa}%
  \BibitemOpen
  \bibfield  {author} {\bibinfo {author} {\bibfnamefont {O.}~\bibnamefont
  {Gazzano}}, \bibinfo {author} {\bibfnamefont {S.~M.}\ \bibnamefont
  {de~Vasconcellos}}, \bibinfo {author} {\bibfnamefont {C.}~\bibnamefont
  {Arnold}}, \bibinfo {author} {\bibfnamefont {A.}~\bibnamefont {Nowak}},
  \bibinfo {author} {\bibfnamefont {E.}~\bibnamefont {Galopin}}, \bibinfo
  {author} {\bibfnamefont {I.}~\bibnamefont {Sagnes}}, \bibinfo {author}
  {\bibfnamefont {L.}~\bibnamefont {Lanco}}, \bibinfo {author} {\bibfnamefont
  {A.}~\bibnamefont {Lema{\^\i}tre}}, \ and\ \bibinfo {author} {\bibfnamefont
  {P.}~\bibnamefont {Senellart}},\ }\href@noop {} {\bibfield  {journal}
  {\bibinfo  {journal} {Nature Communications}\ }\textbf {\bibinfo {volume}
  {4}},\ \bibinfo {pages} {1425} (\bibinfo {year} {2013})}\BibitemShut
  {NoStop}%
\bibitem [{\citenamefont {Haylock}\ \emph {et~al.}(2016)\citenamefont
  {Haylock}, \citenamefont {Lenzini}, \citenamefont {Kasture}, \citenamefont
  {Fisher}, \citenamefont {Streed},\ and\ \citenamefont {Lobino}}]{Ben:23}%
  \BibitemOpen
  \bibfield  {author} {\bibinfo {author} {\bibfnamefont {B.}~\bibnamefont
  {Haylock}}, \bibinfo {author} {\bibfnamefont {F.}~\bibnamefont {Lenzini}},
  \bibinfo {author} {\bibfnamefont {S.}~\bibnamefont {Kasture}}, \bibinfo
  {author} {\bibfnamefont {P.}~\bibnamefont {Fisher}}, \bibinfo {author}
  {\bibfnamefont {E.~W.}\ \bibnamefont {Streed}}, \ and\ \bibinfo {author}
  {\bibfnamefont {M.}~\bibnamefont {Lobino}},\ }\href@noop {} {\bibfield
  {journal} {\bibinfo  {journal} {Rev. Sci. Instrum.}\ }\textbf {\bibinfo
  {volume} {87}},\ \bibinfo {pages} {054709} (\bibinfo {year}
  {2016})}\BibitemShut {NoStop}%
\bibitem [{\citenamefont {Aaronson}\ and\ \citenamefont
  {Arkhipov}(2011)}]{AAronson:18}%
  \BibitemOpen
  \bibfield  {author} {\bibinfo {author} {\bibfnamefont {S.}~\bibnamefont
  {Aaronson}}\ and\ \bibinfo {author} {\bibfnamefont {A.}~\bibnamefont
  {Arkhipov}},\ }\href@noop {} {\bibfield  {journal} {\bibinfo  {journal}
  {Proc. ACM Symposium on Theory of Computing, San Jose, CA}\ ,\ \bibinfo
  {pages} {333}} (\bibinfo {year} {2011})}\BibitemShut {NoStop}%
\bibitem [{\citenamefont {Aaronson}(2011)}]{AAronson:19}%
  \BibitemOpen
  \bibfield  {author} {\bibinfo {author} {\bibfnamefont {S.}~\bibnamefont
  {Aaronson}},\ }\href@noop {} {\bibfield  {journal} {\bibinfo  {journal}
  {Proceedings of the Royal Society A: Mathematical, Physical and Engineering
  Science}\ }\textbf {\bibinfo {volume} {467}},\ \bibinfo {pages} {3393}
  (\bibinfo {year} {2011})}\BibitemShut {NoStop}%
\bibitem [{\citenamefont {Shen}, \citenamefont {Zhang},\ and\ \citenamefont
  {Duan}(2014)}]{Shen:20}%
  \BibitemOpen
  \bibfield  {author} {\bibinfo {author} {\bibfnamefont {C.}~\bibnamefont
  {Shen}}, \bibinfo {author} {\bibfnamefont {Z.}~\bibnamefont {Zhang}}, \ and\
  \bibinfo {author} {\bibfnamefont {L.-M.}\ \bibnamefont {Duan}},\ }\href@noop
  {} {\bibfield  {journal} {\bibinfo  {journal} {Phys. Rev. Lett.}\ }\textbf
  {\bibinfo {volume} {112}},\ \bibinfo {pages} {050504} (\bibinfo {year}
  {2014})}\BibitemShut {NoStop}%
\bibitem [{\citenamefont {Lund}\ \emph {et~al.}(2014)\citenamefont {Lund},
  \citenamefont {Laing}, \citenamefont {Rahimi-Keshari}, \citenamefont
  {Rudolph}, \citenamefont {O'Brien},\ and\ \citenamefont {Ralph}}]{Lund:21}%
  \BibitemOpen
  \bibfield  {author} {\bibinfo {author} {\bibfnamefont {A.~P.}\ \bibnamefont
  {Lund}}, \bibinfo {author} {\bibfnamefont {A.}~\bibnamefont {Laing}},
  \bibinfo {author} {\bibfnamefont {S.}~\bibnamefont {Rahimi-Keshari}},
  \bibinfo {author} {\bibfnamefont {T.}~\bibnamefont {Rudolph}}, \bibinfo
  {author} {\bibfnamefont {J.}~\bibnamefont {O'Brien}}, \ and\ \bibinfo
  {author} {\bibfnamefont {T.}~\bibnamefont {Ralph}},\ }\href@noop {}
  {\bibfield  {journal} {\bibinfo  {journal} {Phys. Rev. Lett.}\ }\textbf
  {\bibinfo {volume} {113}},\ \bibinfo {pages} {100502} (\bibinfo {year}
  {2014})}\BibitemShut {NoStop}%
\bibitem [{\citenamefont {Huh}\ \emph {et~al.}(2015)\citenamefont {Huh},
  \citenamefont {Guerreschi}, \citenamefont {Peropadre}, \citenamefont
  {McClean},\ and\ \citenamefont {Aspuru-Guzik}}]{Huh:22}%
  \BibitemOpen
  \bibfield  {author} {\bibinfo {author} {\bibfnamefont {J.}~\bibnamefont
  {Huh}}, \bibinfo {author} {\bibfnamefont {G.~G.}\ \bibnamefont {Guerreschi}},
  \bibinfo {author} {\bibfnamefont {B.}~\bibnamefont {Peropadre}}, \bibinfo
  {author} {\bibfnamefont {J.~R.}\ \bibnamefont {McClean}}, \ and\ \bibinfo
  {author} {\bibfnamefont {A.}~\bibnamefont {Aspuru-Guzik}},\ }\href@noop {}
  {\bibfield  {journal} {\bibinfo  {journal} {Nat. Photon}\ }\textbf {\bibinfo
  {volume} {9}},\ \bibinfo {pages} {615} (\bibinfo {year} {2015})}\BibitemShut
  {NoStop}%
\bibitem [{\citenamefont {Latmiral}, \citenamefont {Spagnolo},\ and\
  \citenamefont {Sciarrino}(2016)}]{QSup:Latmiral16}%
  \BibitemOpen
  \bibfield  {author} {\bibinfo {author} {\bibfnamefont {L.}~\bibnamefont
  {Latmiral}}, \bibinfo {author} {\bibfnamefont {N.}~\bibnamefont {Spagnolo}},
  \ and\ \bibinfo {author} {\bibfnamefont {F.}~\bibnamefont {Sciarrino}},\
  }\href@noop {} {\bibfield  {journal} {\bibinfo  {journal} {arXiv:1610.02279}\
  } (\bibinfo {year} {2016})}\BibitemShut {NoStop}%
\end{thebibliography}%

% Full bibliography added automatically for Optics Letters submissions
% Note that this extra page will not count against page length

%Manual citation list
%\begin{thebibliography}{1}
%\bibitem{Zhang:14}
%Y.~Zhang, S.~Qiao, L.~Sun, Q.~W. Shi, W.~Huang, %L.~Li, and Z.~Yang,
% \enquote{Photoinduced active terahertz metamaterials with nanostructured
%vanadium dioxide film deposited by sol-gel method,} Opt. Express \textbf{22},
%11070--11078 (2014).
%\end{thebibliography}
\end{document}